\documentclass[preprint,review,12pt]{elsarticle}

\usepackage{amsmath,amsfonts,bm}

\newcommand\vf{{\bm{f}}}
\newcommand\vg{{\bm{g}}}

\newcommand\vr{{\bm{r}}}

\newcommand\vx{{\bm{x}}}
\newcommand\vy{{\bm{y}}}

\newcommand\mH{{\bm{H}}}
\newcommand\mI{{\bm{I}}}

\newcommand\mU{{\bm{U}}}

\newcommand\mZ{{\bm{Z}}}

\usepackage{natbib}
\usepackage{amsmath, amssymb, mathtools, bm, physics}
\usepackage{amssymb, mathtools, physics}
\usepackage{graphicx, subcaption, rotating}      
\usepackage{multirow, makecell, booktabs}        
\usepackage{geometry, setspace, float}           
\usepackage{lineno}                              
\usepackage{appendix}                            
\usepackage{microtype}                           
\usepackage{xcolor}
\usepackage{hyperref}
\usepackage{url}
\usepackage{caption} 
\usepackage[group-separator={,}]{siunitx} 
\newcommand{\mc}{\mathcal}
\newcommand{\mr}{\mathrm}
\newcommand{\mbb}{\mathbb}

\newcommand{\EQ}{\begin{equation}}
\newcommand{\EN}{\end{equation}}
\newcommand{\EQA}{\begin{eqnarray}}
\newcommand{\ENA}{\end{eqnarray}}

\newcommand{\lrr}[1]{\left(#1\right)}
\newcommand{\lrs}[1]{\left[#1\right]}
\newcommand{\lrc}[1]{\left\{#1\right\}}

\newcommand{\lrN}[1]{\left\Vert#1\right\Vert}

\newcommand\Rey{\mbox{\textit{Re}}}  
\newcommand\etal{\mbox{\textit{et al.}}}

\journal{}

\biboptions{sort&compress}

\begin{document}

\graphicspath{{figures/}}

\begin{frontmatter}

\title{Data-driven quantum Koopman method for simulating nonlinear dynamics}

\author[fir]{Baoyang Zhang}

\author[fir]{Zhen Lu\corref{cor1}}
\ead{zhen.lu@pku.edu.cn}

\author[fir,sec]{Yaomin Zhao}

\author[fir,sec]{Yue Yang\corref{cor1}}
\ead{yyg@pku.edu.cn}

\cortext[cor1]{Corresponding author.}

\address[fir]{State Key Laboratory for Turbulent and Complex Systems, School of Mechanics and Engineering Science, Peking University, Beijing 100871, China}
\address[sec]{HEDPS-CAPT, Peking University, Beijing 100871, China}

\begin{abstract}

Quantum computation offers potential exponential speedups for simulating certain physical systems, but its application to nonlinear dynamics is inherently constrained by the requirement of unitary evolution. 
We propose the quantum Koopman method (QKM), a data-driven framework that bridges this gap through transforming nonlinear dynamics into linear unitary evolution in higher-dimensional observable spaces.
Leveraging the Koopman operator theory to achieve a global linearization, our approach maps system states into a hierarchy of Hilbert spaces using a deep autoencoder. 
Within the linearized embedding spaces, the state representation is decomposed into modulus and phase components, and the evolution is governed by a set of unitary Koopman operators that act exclusively on the phase. 
These operators are constructed from diagonal Hamiltonians with coefficients learned from data, a structure designed for efficient implementation on quantum hardware. 
This architecture enables direct multi-step prediction, and the operator's computational complexity scales logarithmically with the observable space dimension. 
The QKM is validated across diverse nonlinear systems. Its predictions maintain relative errors below 6\% for reaction-diffusion systems and shear flows, and capture key statistics in 2D turbulence. 
This work establishes a practical pathway for quantum-accelerated simulation of nonlinear phenomena, exploring a framework built on the synergy between deep learning for global linearization and quantum algorithms for unitary dynamics evolution. 

\end{abstract}

\end{frontmatter}

\section{Introduction}

Quantum computing holds the promise of exponential computational speedups for certain problems~\cite{Feynman1982Simulating}, offering transformative potential for scientific simulation and optimization tasks~\cite{Daley2022Practical}. 
However, the evolution of quantum states is governed by unitary operators, a process that is fundamentally linear and reversible~\cite{nielsen2010}.
This inherent linearity creates a fundamental mismatch with the nonlinear dynamics that characterize most physical phenomena of practical interest, including fluid dynamics~\cite{Givi2020Quantum,Todorova2020Quantum,Fukagata2022quantum} and chemical reactions~\cite{Lu2017Analysis,Akiba2023Carleman,Lu2024Quantum}. 
Bridging this gap represents a central challenge for realizing quantum advantages in real-world applications~\cite{Succi2023Quantum,Meng2024Challenges,Tennie2025Quantum}. 
Moreover, any viable solution must demonstrate not only theoretical compatibility but also computational efficiency that effectively leverages quantum resources to achieve speedups over classical approaches~\cite{Hoefler2023Disentangling,Aaronson2025Future}.

To address the challenges of nonlinearity and the limitations of noisy intermediate-scale quantum (NISQ)~\cite{Preskill2018Quantum} hardware, a common strategy is the hybrid quantum-classical framework~\cite{Gaitan2020Finding,Budinski2022Quantum,Pfeffer2022Hybrid,Pfeffer2023Reducedorder,Ahmed2024Prediction,Bharadwaj2023Hybrid,Bharadwaj2024QFlowS,Bharadwaj2025Compact,Liu2023Quantum,Chen2022Quantum,Chen2024Enabling,Ye2024hybrid,Sedykh2024Hybrid,Song2025Incompressible,Choi2025variational,Jaksch2023Variational}. 
In these frameworks, quantum computers execute subroutines with efficient quantum algorithms~\cite{harrow2009quantum,Costa2022Optimal,Jaksch2023Variational,Asztalos2024Reducedorder}, while classical computers handle operations unsuitable for quantum implementation, particularly those involving strong nonlinearity~\cite{Bharadwaj2023Hybrid,Ye2024hybrid}. 
However, the iterative quantum-classical data exchange introduces a critical bottleneck~\cite{Chen2022Quantum}. 
When repeated at every time step, this overhead can overwhelm any potential quantum speedup, creating a significant barrier to practical speedup~\cite{Aaronson2015Read}. 
These limitations motivate alternative approaches that circumvent iterative quantum-classical cycles, particularly through linearization of nonlinear dynamics~\cite{jin2023time,jin2024BCM}, enabling non-iterative quantum computing.

One pathway expands nonlinear systems into linear approximations amenable to quantum algorithms. 
Carleman linearization embeds polynomial nonlinear ordinary differential equations (ODEs) into infinite-dimensional linear systems~\cite{carleman1932,kowalski1991nonlinear}.
Liu \etal~\cite{liu2021efficient} implemented Carleman linearization to enable quantum solution methods for previously intractable nonlinear systems. 
Subsequent works have applied Carleman linearization to computational fluid dynamics~\cite{Sanavio2024Three,Sanavio2025Carleman,Gonzalez-Conde2025Quantum} and chemical reaction systems~\cite{Akiba2023Carleman}.
Similarly, Xue \etal~\cite{Xue2021Quantum} applied homotopy perturbation methods to quadratic nonlinear ODEs and extended the homotopy analysis method~\cite{liao1992proposed,Liao2004homotopy,Liao2009Notes} to nonlinear partial differential equations (PDEs)~\cite{Xue2025Quantum}. 
However, these approaches face convergence challenges, restricting validity to short time intervals, weak nonlinearity, or specialized cases. 

An alternative pathway converts nonlinear dynamics into exact linear equations via mathematical transformations, preserving the complete nonlinear dynamics. 
The Liouville~\cite{jin2023time,Succi2024Ensemble} and Fokker-Planck~\cite{Gourianov2025Tensor,Tennie2024Solving} equations describe the deterministic and stochastic dynamics through linear PDEs on probability density function (PDF), respectively.
These linear PDEs can be solved using quantum algorithms, such as linear combination of Hamiltonian simulation~\cite{An2023Linear,An2025Quantum}, Schr\"{o}dingerization~\cite{Jin2023Quantum,jin2024quantumPRL,jin2025quantum}, and block-encoding~\cite{Tennie2024Solving,Brearley2024Quantum}.
The Koopman-von Neumann approach~\cite{joseph2020koopman,Novikau2025Quantum} introduces complex wave functions whose squared magnitude yields the PDF.
The wave function evolves via a Schr\"{o}dinger-like equation with a Hermitian Hamiltonian operator, enabling direct quantum simulation. 
While these transformation methods avoid approximation, they suffer from exponential scaling in system dimensionality~\cite{Jin2022Time}, requiring exponentially large quantum resources as the number of variables increases.

Note certain specialized transformations can linearize nonlinear dynamics without increasing dimensionality. 
Meng and Yang~\cite{meng2023quantum,meng2024quantum} utilized the generalized Madelung transformation~\cite{Madelung1927Quantentheorie,chern2016schrodinger,zylberman2022quantum,Yang2023Applications} to map fluid governing equations to the hydrodynamic Schr\"{o}dinger equation, establishing a framework for quantum computing of fluid dynamics. 
This framework has been implemented and validated on real superconducting quantum computers for unsteady flows~\cite{Meng2024Simulatinga}. 
Nevertheless, finding such global linearization transformations for general nonlinear dynamical systems remains challenging and is not straightforward for arbitrary systems.

The Koopman theory~\cite{koopman1931hamiltonian,mezic2013analysis,Brunton2022Modern} linearizes nonlinear dynamics by lifting state evolution to an infinite-dimensional observable space. 
The central challenge is identifying a finite-dimensional invariant subspace that captures the essential dynamics. 
Deep learning has emerged as an approach to find the observable and operator~\cite{lusch2018deep,gin2021deep,Eivazi2021Recurrent,Xiong2023KoopmanLab,xiong2024koopman,Yu2024Parametric,yu2025koopman,Zhou2025Neural}, with successful applications to fluid dynamics~\cite{lusch2018deep,gin2021deep}, turbulent flows~\cite{Eivazi2021Recurrent, xiong2024koopman}, and flame dynamics~\cite{Yu2024Parametric,yu2025koopman}.
However, these methods generate non-unitary representations incompatible with quantum simulation.
A quantum implementation of Koopman theory requires both finite-dimensionality and unitarity~\cite{giannakis2021quantum}. 
Theoretical analysis establishes that measure-preserving, ergodic dynamical systems yield unitary Koopman operators, which is validated on dynamics of harmonic oscillators~\cite{giannakis2022embedding}. 
However, the demanding condition excludes general nonlinear systems.

We propose the quantum Koopman method (QKM), a data-driven framework bridging quantum Koopman theory and simulations of nonlinear dynamics, including reaction-diffusion systems and fluid dynamics. 
QKM employs a hierarchical decomposition~\cite{li2025transformer} to approximate dynamical systems using multiple learned ergodic rotations, each operating on finite-dimensional Hilbert spaces.
By separating modulus and phase components and restricting temporal evolution to the phase space, QKM learns unitary operators while maintaining the flexibility for complex nonlinear phenomena.
The resulting factorizable Hamiltonian enables quantum circuit implementation using only single-qubit rotations~\cite{giannakis2022embedding}, ensuring hardware compatibility and exponential speedup over classical Koopman implementations.
We validate QKM on reaction-diffusion systems, shear flows, and 2D turbulence, demonstrating its capability to capture essential dynamics.

The paper is organized as follows. Section~\ref{sec:theory} establishes the Koopman theory and its quantum embedding for classical dynamical systems. Section~\ref{sec:qkm} details the QKM framework and its implementation. 
Section~\ref{sec:results} presents comprehensive numerical validation across three canonical cases: the reaction-diffusion system, shear flow, and two-dimensional (2D) turbulence. 
Section~\ref{sec:conclusion} presents the conclusions. 

\section{Koopman theory}\label{sec:theory}

\subsection{Koopman operator}

The Koopman theory linearizes nonlinear dynamics by lifting evolution from a finite-dimensional state space to an infinite-dimensional space of observables~\cite{koopman1931hamiltonian}. 
Consider the dynamical system
\EQ\label{eq:dyn}
    \dv{}{t}\vx\lrr{t} = \vg\lrs{\vx\lrr{t}}, 
\EN
where $\vx \in \mc{X}$ is the system state on a manifold $\mc{X} \subset \mbb{R}^d$ of $d$-dimensional real space $\mbb{R}^d$, and $\vg$ can be a nonlinear vector field.  
The solution of Eq.~\eqref{eq:dyn} defines a flow map $\Phi^t \colon \mc{X} \to \mc{X}$, such that $\vx\lrr{t} = \Phi^t\lrs{\vx\lrr{0}}$ for initial condition $\vx\lrr{0}$.

Instead of evolving state $\vx$ directly, the Koopman theory considers the evolution of observables $f\colon \mc{X} \to \mbb{C}$, where $\mbb{C}$ denotes the complex space. 
The complex-valued observables naturally encode both modulus and phase information, enabling the construction of unitary evolution operators required for quantum implementation. 
The Koopman operator $\mU^t$ acts on the space of observables as 
\EQ\label{eq:def}
    \mU^t f\lrr{\vx} = f\circ\Phi^t\lrr{\vx}, 
\EN
where $\circ$ denotes function composition, $f\circ\Phi^t\lrr{\vx} = f\lrs{\Phi^t\lrr{\vx}}$.
Although the flow map $\Phi^t$ may be nonlinear, the Koopman operator $\mU^t$ is linear by construction. 
The fundamental challenge lies in identifying finite-dimensional subspaces of observables that are approximately invariant under the Koopman operator and capture essential system dynamics. 

\subsection{Finite-dimensional unitary representation}

Quantum computing implementation requires the Koopman operator to be both finite-dimensional and unitary~\cite{giannakis2021quantum,giannakis2022embedding}. 
A model demonstrating these properties is the ergodic rotation on the $d$-dimensional torus $\mathbb{T}^d$.
This system admits a unitary Koopman operator $\mU^t = e^{i\mH t}$ on an appropriate Hilbert space of observables, with the Hermitian matrix $\mH$~\cite{giannakis2021quantum}. 
This structure directly parallels quantum mechanics, with observables $f\lrr{\mc{X}}$ evolving according to Schr\"odinger-like equations.

Projecting $\mH$ onto the Hilbert space $\mbb{C}^{2^n}$ yields an approximation $\hat{\mH}$ and the associated Koopman operator~\cite{giannakis2022embedding} 
\EQ\label{eq:Unt}
    \hat{\mU}^t = e^{i\hat{\mH}t}.
\EN
For systems with pure point spectrum, $\hat{\mH}$ can be decomposed as~\cite{welch2014efficient}
\begin{equation}\label{eq:Hn}
    \hat{\mH} = -\frac{1}{2} \sum_{k=1}^{n} \alpha_k \lrs{ \mI^{\otimes (k-1)} \otimes \mZ \otimes \mI^{\otimes (n-k)} },
\end{equation}
where 
\begin{equation*}
    \mZ = \begin{pmatrix} 1 & 0 \\ 0 & -1 \end{pmatrix} 
    \quad \mr{and} \quad 
    \mI = \begin{pmatrix} 1 & 0 \\ 0 & 1 \end{pmatrix}
\end{equation*}
are the Pauli-Z and identity matrices, respectively, $\lrc{\alpha_k}_{k=1}^n$ are real coefficients, and $\otimes$ denotes tensor product.
Substituting Eq.~\eqref{eq:Hn} into Eq.~\eqref{eq:Unt} yields the factorized Koopman operator 
\EQ\label{eq:Kop}
    \hat{\mU}^t = \bigotimes_{k=1}^n R_z(\alpha_k t),
\EN
where $R_z(\alpha_k t) = e^{-i \mZ \alpha_k t /2}$ represents the single-qubit $z$-rotation~\cite{nielsen2010}. 
This tensor product decomposes the evolution in $\mbb{C}^{2^n}$ into $n$ independent 2D rotations, enabling efficient quantum circuit implementation with $\mc{O}\lrr{n}$ complexity for $N=2^n$ dimensional systems~\cite{giannakis2022embedding}. 

However, this unitary representation applies only to systems that are topologically conjugate to ergodic rotations on a torus, a highly demanding condition. 
Most nonlinear dynamics of practical interest, especially those exhibiting chaotic or dissipative behaviors, do not meet this requirement. 
Moreover, identifying appropriate observables remains a fundamental challenge.
These limitations motivate employing the data-driven approach~\cite{Brunton2022Modern} to learn both the observable mapping $f\lrr{\mc{X}}$ and corresponding evolution operator $\hat{\mU}^t$.

\section{Data-driven quantum Koopman method}\label{sec:qkm}

We propose a data-driven QKM that learns both the observable mappings and unitary evolution operator from data. 
Our approach employs a subsystem decomposition to map the state space into a hierarchy of finite-dimensional Hilbert spaces. 
Each subsystem is governed by a learned unitary Koopman operator with the tensor product structure of Eq.~\eqref{eq:Kop}, enabling efficient quantum implementation.
This hierarchical framework extends conventional Koopman theory by employing multiple learned operators that collectively approximate complex dynamical systems. 

\subsection{Quantum Koopman method}

\begin{figure}[!ht]
    \centering
    \includegraphics[width=1.0\linewidth]{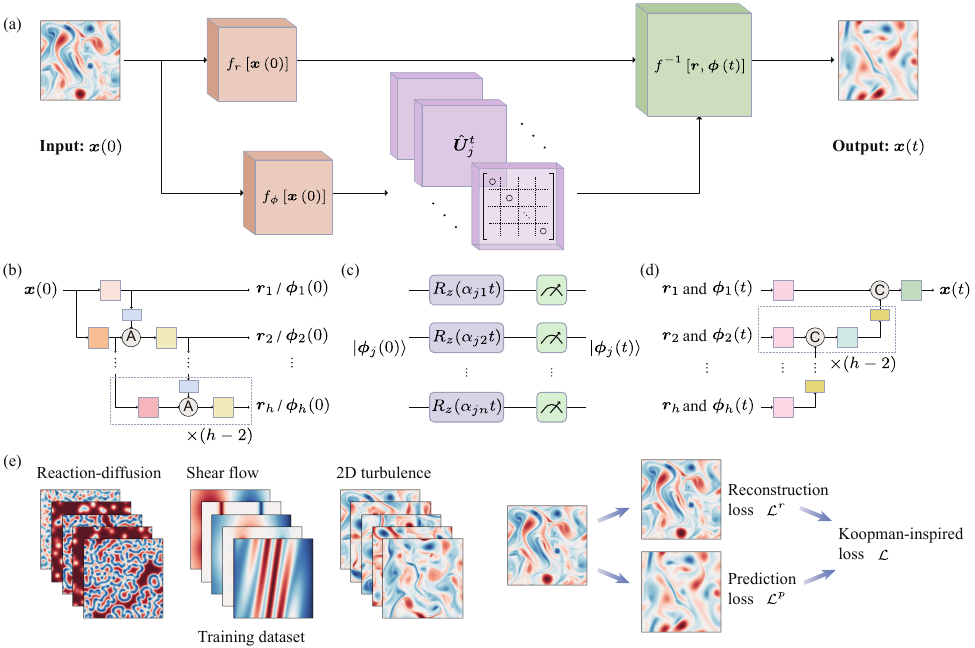}
    \caption{
    Overview of the data-driven QKM.
    (a) Overall framework: the input state $\vx\lrr{0}$ is processed through parallel encoders to obtain modulus components $\vr_j$ and phase components $\bm{\phi}_j$ across $h$ subsystems. 
    Each subsystem's phase undergoes temporal evolution via learned unitary Koopman operators $\hat{\mU}_j^t$ while the modulus components remain invariant.
    The evolved components are then integrated and decoded to produce the predicted output $\vx\lrr{t}$. 
    (b) Observable encoder architecture: both modulus and phase encoders employ identical hierarchical neural network architecture mapping the input state to the respective subsystem components. 
    (c) Quantum circuit for evolution: each unitary Koopman operator $\hat{\mU}_j^t$ is constructed using tensor products of $R_z$ rotation gates with learnable parameters $\{\alpha_{jk}\}_{k=1}^{n_j}$, enabling efficient quantum circuit implementation with $\mc{O}\lrr{n_j}$ complexity. 
    (d) Observable decoder architecture: the decoder network reconstructs the predicted state from the concatenated modulus and evolved phase components across $h$ subsystems. 
    (e) Training framework: the model is trained on diverse dynamical systems including reaction-diffusion, shear flow, and 2D turbulence using a combined loss function comprising reconstruction and prediction losses to jointly optimize the autoencoder and evolution operators.
    }
    \label{fig:net}
\end{figure}

The overall QKM framework is illustrated in Fig.~\ref{fig:net}(a). 
We approximate the infinite-dimensional observable space using $h$ finite-dimensional subsystems in a hierarchical structure.
Each subsystem operates within a $2^{n_j}$-dimensional Hilbert space, induced by a quantum feature map of $n_j$ qubits, where the subscript $j=1, \dots, h$ denote the $j$-th subsystem and $N_j = 2^{n_j}$.

To satisfy the unitary evolution requirement, we decompose the state space mapping into modulus and phase components. 
The observable function maps the state space $\mc{X}$ as
\EQ\label{eq:decomp}
    f\lrr{\bm{x}} = \bm{r} \odot e^{i\bm{\phi}},
\EN
where $\odot$ represents element-wise multiplication, 
$\bm{r} = \bigoplus_{j=1}^h \bm{r}_j$
and
$\bm{\phi} = \bigoplus_{j=1}^h \bm{\phi}_j$
are the concatenation of subsystem modulus $\lrc{\vr_j}_{j=1}^h$ and phase $\lrc{\bm{\phi}_j}_{j=1}^h$, respectively, and $\bigoplus$ denotes vector concatenation. 

As shown in Fig.~\ref{fig:net}(a), the modulus $\vr = f_r \lrr{\vx}$ and phase $\bm{\phi} = f_\phi \lrr{\vx}$ components are obtained via parallel learned mappings, where both encoders $f_r$ and $f_\phi$ employ identical architectures, as detailed in Fig.~\ref{fig:net}(b). 
The temporal evolution operates exclusively on the phase components through a block-diagonal unitary Koopman operator
\EQ\label{eq:Ut}
    \hat{\mU}^t 
    = 
    \begin{pmatrix}
        \hat{\mU}_1^t &         & \\
                      & \ddots  & \\
                      &         & \hat{\mU}_h^t
    \end{pmatrix},
\EN
where each block 
\EQ\label{eq:ujt}
    \hat{\mU}_j^t = \bigotimes_{k=1}^{n_j} R_z(\alpha_{jk} t) \quad \text{for } j = 1,2,\ldots,h
\EN
is the Koopman operator for the $j$-th subsystem with learnable parameters $\lrc{\alpha_{jk}}_{k=1}^{n_j}$.
Each subsystem's phase evolves through $\hat{\mU}_j^t$ composed of $n_j$ quantum gates; at the same time, the modulus components remain unchanged, as illustrated in Fig.~\ref{fig:net}(c). 
Individual subsystem evolution follows
\EQ\label{eq:phaseEvo}
    \vr_j\odot e^{i \bm{\phi}_j\lrr{t}} = \vr_j\odot \hat{\mU}_j^t e^{i \bm{\phi}_j\lrr{0}}
    \quad \mr{for} \quad
    j = 1, 2, \dots, h.
\EN
This encoding of the phase angle enables learning an operator with the unitary constraint.

Finally, all processed subsystems are integrated and the time-evolved state $\vx(t)$ is reconstructed via learned mapping $\vx(t) = f^{-1} \lrs{ \vr, \bm{\phi}\lrr{t} }$, using the architecture in Fig.~\ref{fig:net}(d).
The autoencoder architecture including $f_r$, $f_\phi$, and $f^{-1}$ are detailed in~\ref{app:ae}.
The autoencoder and Koopman operators are jointly optimized via a data-driven approach in Fig.~\ref{fig:net}(e), using a combined reconstruction and prediction loss function. 

The hierarchical encoding creates subsystems with decreasing dimensions. 
For a system state $\vx \in \mbb{R}^d$, the dimension of the $j$-th subspace is 
\EQ\label{eq:Nj}
N_j = 2^{1-j} c d \quad \mr{for} \quad j = 1, 2, \dots, h, 
\EN
where the configurable integer $c$ is the expansion ratio from the state dimension $d$ to the first subsystem dimension $N_1 = c d$ and represents the number of feature map channels.
The total observable dimension $N= \sum_{j=1}^h N_j$ across $h$ subsystem is
\EQ\label{eq:N}
N = (2 - 2^{1-h}) c d.
\EN
The configurable parameters $c$ and $h$ allow flexible control over the approximation fidelity when using $h$ ergodic rotations to approximate general system dynamics.

This framework extends the classical Koopman operator theory in Eq.~\eqref{eq:def} by employing learned unitary operators of subsystems, Eq.~\eqref{eq:ujt}.
Each subsystem is represented as an ergodic rotation on a torus, approximating the original dynamical system by a collection of ergodic rotations. 
Modulus-phase decomposition in Eq.~\eqref{eq:decomp} and phase encoding enable learnable unitary Koopman operators. 
This establishes a unified global linearization framework for efficient quantum simulation of classical nonlinear systems. 

\subsection{Unitary Koopman operators}

The QKM exploits the factorized structure of $\hat{\mU}_j^t$ in Eq.~\eqref{eq:ujt} to achieve efficient simulation. 
For the $j$-th subsystem, the phase component $\bm{\phi}_j$ is an $N_j$-dimensional vector.
We encode $\bm{\phi}_j$ into a quantum state of $n_j$ qubits, where each qubit $\ket{l_m}$ with $m = 1, \cdots, n_j$ represents a computational basis state $\ket{0}$ or $\ket{1}$. 
The computational basis $\ket{l} = \bigotimes_{m=1}^{n_j} \ket{l_m} $ spans the computational domain in the Hilbert space $\mbb{C}^{2^{n_j}}$, with $l = \sum_{m=1}^{n_j} l_m 2^{n_j-m}$~\cite{nielsen2010}.
The quantum state is encoded as 
\EQ
    \ket{\bm{\phi}_j} = \dfrac{1}{\sqrt{N_j}}\sum_{l=1}^{N_j} e^{i \bm{\phi_{jl}}} \ket{l},
\EN
where $\bm{\phi}_{jl}$ denotes the $l$-th component of the phase vector $\bm{\phi}_j$.

For temporal discretized data, QKM learns the operator $\hat{\mU}^{\Delta t} = \bigotimes_{m=1}^{n_j} R_z(\alpha_{jm} \Delta t)$ for one time step $\Delta t$. 
The quantum temporal evolution follows the unitary dynamics
\EQ\label{eq:phiEvo}
    \ket{\bm{\phi}_j(k\Delta t)} = \lrr{\hat{\mU}_j^{\Delta t}}^k \ket{\bm{\phi}_j(0)}
    \quad \mr{for} \quad k = 0, 1, 2, \cdots,
\EN
where $ (\hat{\mU}^{\Delta t})^k = \bigotimes_{m=1}^{n_j} R_z(\alpha_{jm} k \Delta t)$.
This global linearization avoids iterative quantum state measurement and preparation.
The quantum circuit implementation for Eq.~\eqref{eq:phiEvo} is shown in Fig.~\ref{fig:net}(c). 
Each unitary Koopman operator $(\hat{\mU}_j^{\Delta t})^k$ requires only $n_j$ parallel single-qubit $R_z$ gates. 
This circuit structure is hardware-friendly, requiring no entangling gates and enabling straightforward implementation on quantum devices with minimal coherence requirements.
Furthermore, the block-diagonal structure of Eq.~\eqref{eq:Ut} allows fully decoupled simulation across subsystems. 

The quantum evolution framework provides significant computational advantages over classical Koopman operator implementations. 
Classical Koopman implementations require $\mc{O}\lrr{N_j}$ operations per evolution for the $j$-th subsystem due to the high-dimensional matrix-vector operations in the observable space. 
By contrast, QKM achieves $\mc{O}\lrr{\log N_j}$ complexity. 
For the complete system with $h$ subsystems, the total complexity for evolution scales as $\mc{O}\lrr{h(\log c + \log d - h)}$. 
Meanwhile, the classical implementation scales as $\mc{O}(N) = \mc{O}((2-2^{1-h})cd)$. 
The exponential advantage becomes increasingly significant as the observable space dimension grows, enabling quantum simulation of high-dimensional Koopman dynamics that would be computationally prohibitive using classical Koopman implementations.

It is important to distinguish the quantum advantage over classical Koopman implementations from comparisons with classical discretization approaches. 
Classical discretization methods, such as finite difference and finite element approaches, operate directly on the original nonlinear system and typically transform the problem into sparse linear algebra equations with $\mc{O}\lrr{d}$ complexity using specialized solvers.
Furthermore, the classical discretization methods for nonlinear dynamics requires iterations for time marching. 
Meanwhile the global linearization via Koopman operator enables an ``one-shot'' solution for given time. 
The quantum advantage over classical discretization approaches depends critically on the relationship between observable space dimension in Eq.~\eqref{eq:N} and original state dimension $d$. 
The QKM maintains its speedup capability when $c < \mc{O}\lrr{2^d/d}$. It means that the observable space cannot grow exponentially with the state dimension while preserving quantum speedup. 
This requirement is non-trivial and depends on the intrinsic complexity of the nonlinear dynamics and the quality of the Koopman approximation.
For systems where effective Koopman representations exist with moderate observable dimensions, such as those with dominant low-dimensional attractors or separable dynamics, quantum advantage becomes achievable.
However, for strongly nonlinear systems requiring exponentially large observable spaces for accurate representation, the quantum advantage may be compromised. 

\subsection{Koopman-inspired loss functions}

Training the QKM requires joint optimization of the autoencoder and Koopman operator~\cite{lusch2018deep}.
This necessitates a composite loss function with reconstruction and prediction components.
Our training dataset comprises multiple time series trajectories of nonlinear dynamics, each containing $T+1$ consecutive states $\vx_0, \vx_1, \dots, \vx_T$ sampled at discrete time steps with interval $\Delta t$ and $\vx_k = \vx\lrr{k\Delta t}$. 

The reconstruction loss
\EQ\label{eq:loss_r}
    \mc{L}^r = \mbb{E}_{\vx_0} \lrs{\frac{1}{T+1} \sum_{k=0}^T  \dfrac{ \lrN{ f^{-1}\circ f\lrr{\vx_k} - \vx_k}_2^2 }{ \lrN{\vx_k}_2^2 } }
\EN
measures the difference between the auto-encoded state \( f^{-1} \circ f(\vx_k) \) and the ground truth (GT) \( \vx_k \), where $\lrN{\cdot}_2$ denotes the $L_2$-norm. 
It ensures accurate mappings between the states and observables. 
The prediction loss
\EQ\label{eq:loss_p}
    \mc{L}^p = \mbb{E}_{\vx_0} \lrs{\frac{1}{T} \sum_{k=1}^T \dfrac{ \lrN{ f^{-1}\circ \lrr{\hat{\mU}^{\Delta t}}^k f \lrr{\vx_0} - \vx_k }_2^2 }{ \lrN{\vx_k}_2^2 } }
\EN
measures the discrepancy between the predicted evolved state \( f^{-1} \circ (\hat{U}^{\Delta t})^k f(\vx_0) \) and the corresponding GT. 

We implement the supervised learning with data organized as pairs $(\vx_{k}, \Delta k)$, where $\Delta k \in \lrs{0, T}$ represents the time step interval. 
The QKM learns to predict $\vx_{k+\Delta k}$ given the input $(\vx_k, \Delta k)$, minimizing
\begin{equation}\label{eq:loss_e}
    \mathcal{L} = \mathbb{E}_{(\vx_{k}, \Delta k)} \lrs{ \frac{ \lrN{ f^{-1} \circ \lrr{\hat{\mU}^{\Delta t}}^{\Delta k} f \lrr{\vx_k} - \vx_{k+\Delta k} }^2_2}{\lrN{ \vx_{k+\Delta k} }^2_2} }.
\end{equation}
Each pair $(\vx_k, \Delta k)$ is indexed by $(k,\Delta k) \in \mathcal{I}$, where
\begin{equation}
    \mathcal{I} = \underbrace{\{(\ell, 0) \mid \ell \in \{0,\dots,T\}\}}_{\text{Zero-step transitions}}
    \cup \underbrace{\{(0, \ell) \mid \ell \in \{1,\dots,T\}\}}_{\text{Initial rollouts}}.
\end{equation}
The zero-step transitions verify identity mappings when $\Delta k = 0$, while initial rollouts evaluate predictions.

Although equivalent to the linear combination of Eqs.~\eqref{eq:loss_r} and \eqref{eq:loss_p}, Eq.~\eqref{eq:loss_e} provides superior memory efficiency.
The linear combination processes entire $(T+1)$-step sequences as single samples, whereas Eq.~\eqref{eq:loss_e} treats time step pairs as individual samples. 
This $T$-fold memory reduction enables training computationally intensive tasks of complex systems like 2D turbulence. 

\section{Results}\label{sec:results}

We performed numerical experiments on three representative nonlinear systems: reaction-diffusion system, shear flow, and 2D turbulence, spanning diverse spatio-temporal characteristics.
For each system, multiple trajectories were generated using different initial conditions~\cite{ohana2024well,kochkov2021machine}.
The datasets were partitioned into training, validation, and test sets in an 8:1:1 ratio. 
Model parameters were optimized using the training set, employing a cosine decay learning rate schedule with logarithmic warm-up, while the validation set guided early stopping. 
Final performance was assessed on the test set. 

All experiments used a trajectory of $T = 60$.
Each trajectory contributes 60 multi-step rollouts and 61 zero-step transitions.
The experiments were conducted using four NVIDIA A800 GPUs, each has 80 GB memory. 
We employed a micro-batch size of 240 per GPU with two gradient accumulation steps, yielding an effective global batch size of 1920. 
This configuration balanced memory efficiency with computational performance. 

The hyperparameters $h$ and $c$ were set as follows: 
$h=4$ and $c=16$, denoted as h4c16, for the reaction-diffusion system, 
and $h=5$ and $c=16$, denoted as h5c16, for both shear flow and 2D turbulence.
Prediction accuracy was quantified using the relative $L_2$ error
\begin{equation}\label{eq:L2e}
    \mr{relative}\;L_2\text{-}\epsilon
    =
    \dfrac{\lrN{ f^{-1} \circ \lrr{\hat{\mU}^{\Delta t}}^k f (\vx_0) - \vx_{k} }^2_2}{\lrN{ \vx_{k} }^2_2}.
\end{equation}

\subsection{Reaction-diffusion system}

The Gray-Scott equation~\cite{gray1983autocatalytic} describes a reaction-diffusion system
\begin{align}
    \frac{\partial Y_A}{\partial t} &= D_A \nabla^2 Y_A - Y_A Y_B^2 + F(1 - Y_A), \label{eq:gray_scott_A} \\
    \frac{\partial Y_B}{\partial t} &= D_B \nabla^2 Y_B + Y_A Y_B^2 - (F + K)Y_B, \label{eq:gray_scott_B}
\end{align}
where $Y_A(x, y, t)$ and $Y_B(x, y, t)$ denote the concentrations of two species with diffusion coefficients $D_A = 2.1 \times 10^{-5}$ and  $D_B = 1.1 \times 10^{-5}$, respectively, 
$F$ is the feed rate of species $A$, and $K$ is the decay rate of species $B$. 
Varying parameters $(F, K)$ produces diverse patterns from random initial conditions. 
We define the dimensionless time $t^* = t / \tau$ with $\tau = 1/F$.

We employed the simulation data from the Well datasets~\cite{ohana2024well}. 
The computational domain spans $x,y \in [-1,1]$ with periodic boundary condition and $128^2$ spatial resolution. 
Data are stored with $\Delta t = 10$.
The original dataset contains six parameters sets $(F, K)$, with 200 random initial conditions per set.
We extracted the six consecutive 61-step segments from each 1001-step solution to capture unsteady dynamics. 
This yields 1200 61-step trajectories per parameter set.

\begin{figure}[!ht]
    \centering
    \includegraphics[width=1.0\linewidth]{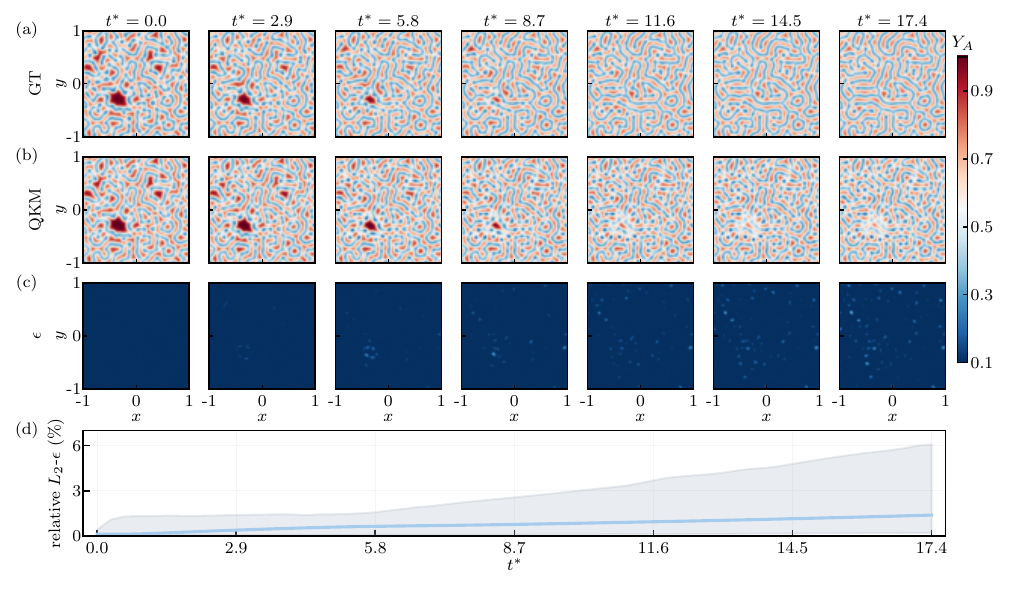}
    \caption{
    Species concentration $Y_A$ over $T = 60$ time steps for $(F, K) = (0.029, 0.057)$. 
    (a) GT, (b) QKM predictions, (c) Pointwise absolute error $\epsilon(x,y;t^{*}) = \bigl|Y_A^{\mathrm{QKM}}(x,y;t^{*}) - Y_A^{\mathrm{GT}}(x,y;t^{*})\bigr|$, and (d) Relative $L_2$ error over time for this case (solid line) with error band representing the 10th to 90th percentile of values across all test reaction-diffusion cases.  
    }
    \label{fig:gray_contour}
\end{figure}

\begin{figure}[!ht]
    \centering
    \includegraphics[width=1.0\linewidth]{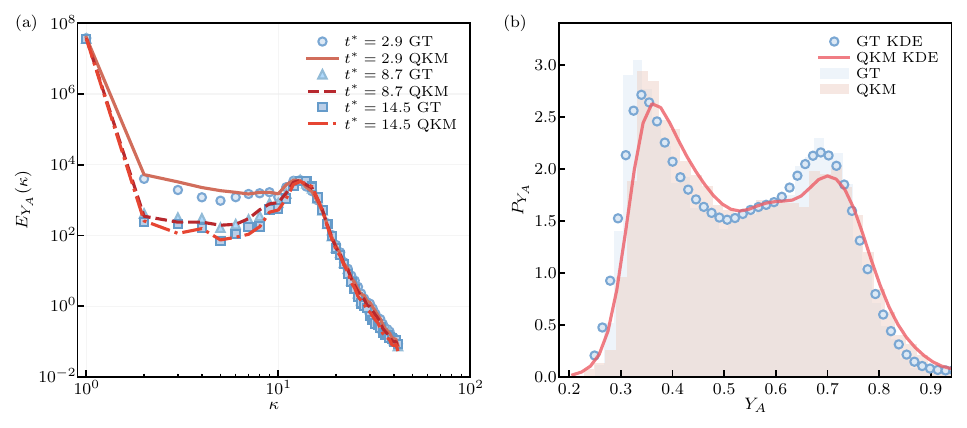}
    \caption{
    Statistics of \(Y_{A}\) for the reaction-diffusion system. 
    (a) Energy spectra \(E_{Y_A}(\kappa)\) comparing GT (blue markers) and QKM at \(t^{*}=2.9\) (solid red), \(t^{*}=8.7\) (dashed red), and \(t^{*}=14.5\) (dash–dot red).
    (b) PDF \(P_{Y_{A}}\) at \(t^{*}=2.9\) comparing histograms (bars) and KDE (lines) for GT (blue) and QKM (red).
    }
    \label{fig:gray_stat}
\end{figure}

For $(F, K) = (0.029, 0.057)$, which produces characteristic maze-like patterns~\cite{pearson1993complex}, Figure~\ref{fig:gray_contour} shows QKM's prediction of $Y_A$ over 60 time steps, comparing with GT. 
The GT contours in Fig.~\ref{fig:gray_contour}(a) reveal maze-like pattern evolution with notable decay of the high-concentration region.
The QKM predictions in Fig.~\ref{fig:gray_contour}(b) accurately reproduce these patterns. 
The pointwise absolute error fields in Fig.~\ref{fig:gray_contour}(c) remain spatially localized with low magnitudes, while Fig.~\ref{fig:gray_contour}(d) shows relative $L_2$ error staying below 2\% for this case throughout the simulation.
Across all test cases on the reaction-diffusion system, QKM demonstrates robust performance, with the 10th to 90th percentile range of the relative $L_2$ error values spanning approximately 0\% to 6\%, as shown in Fig.~\ref{fig:gray_stat}(d). 
Note that unlike conventional numerical integration schemes that require iterative time-stepping, QKM provides direct one-shot predictions by applying the learned linear operator to advance from initial condition to the target.
Test on QKM generalization beyond 60 steps is presented in \ref{app:extrapolation}.

Figure~\ref{fig:gray_stat}(a) compares energy spectra $E_{Y_A}(\kappa)$ between GT and QKM at three time instances, with $E_{Y_A}\equiv Y_A^2/2$ and the wavenumber $\kappa$. 
The spectral evolution from $t^* = 2.9$ to 14.5 shows the dissolution of large-scale high-concentration structure observed in Fig.~\ref{fig:gray_contour} and a ``frozen'' behavior at high wavenumbers, where rapid dissipation drives stabilization. 
QKM demonstrates excellent agreement across the entire wavenumber range, accurately capturing both large-scale and small-scale energy distributions. 

Figure~\ref{fig:gray_stat}(b) shows the species concentration PDF $P_{Y_A}$ via both histogram and kernel density estimation (KDE). 
The pronounced bimodal distribution with peaks around $Y_A = 0.3$ and 0.7 indicates two distinct concentration states corresponding to the maze-like pattern structure. 
QKM reproduces this bimodal character, demonstrating its capability to preserve the statistical properties of the concentration field that emerge from the underlying reaction-diffusion dynamics.

\subsection{Shear flow}\label{sec:shear}

The shear flow is governed by the incompressible Navier-Stokes (NS) equation
\begin{align}\label{eq:NS}
    \nabla \cdot \bm{u} & = 0, \\
    \frac{\partial \bm{u}}{\partial t} + \bm{u} \cdot \nabla \bm{u} & = -\nabla p + \frac{1}{\Rey} \nabla^2 \bm{u} + \vf,
\end{align}
where $\bm{u}$ is the velocity, $p$ is the pressure, $\Rey$ is the Reynolds number, and $\vf$ is the body force term, which is set to 0 for the shear flow. 
The initial conditions consist of $m_s$ uniformly spaced tanh shear profiles with sinusoidal perturbations characterized by the blob mode number $m_b$ and length scale $l_s$~\cite{ohana2024well}. 
We define the dimensionless time $t^* = t \langle |\omega_0|_{\mr{max}} \rangle$, where $\langle|\omega_0|_{\mr{max}}\rangle$ is the ensemble-averaged maximum magnitude of vorticity $\omega = \nabla \times \bm{u}$ at $t=0$.

We used simulation data from the Well datasets~\cite{ohana2024well, burns2020dedalus}.
The computational domain spans $x \in [0,1]$ and $y \in [0,1]$, with a $128^2$ spatial resolution, downsampled from half of the original domain $x \in [0,1]$ and $y \in [-1,1]$ with $256\times 512$. 
The cases have $\Rey = 50000$, characterized by shear mode $m_s \in \{1, 2\}$, blob mode $m_b \in \{2, 3, 4, 5\}$, and length scale $l_s \in \{0.25, 0.5, 1.0, 2.0, 4.0\}$. Loop over the parameters yields 40 initial conditions. 
Each case spans $t \in [0, 20]$ with $\Delta t = 0.1$.
We extracted the three consecutive 61-step segments from each 201-step solution, yielding 120 trajectories in total.

\begin{figure}[!ht]
    \centering
    \includegraphics[width=1.0\linewidth]{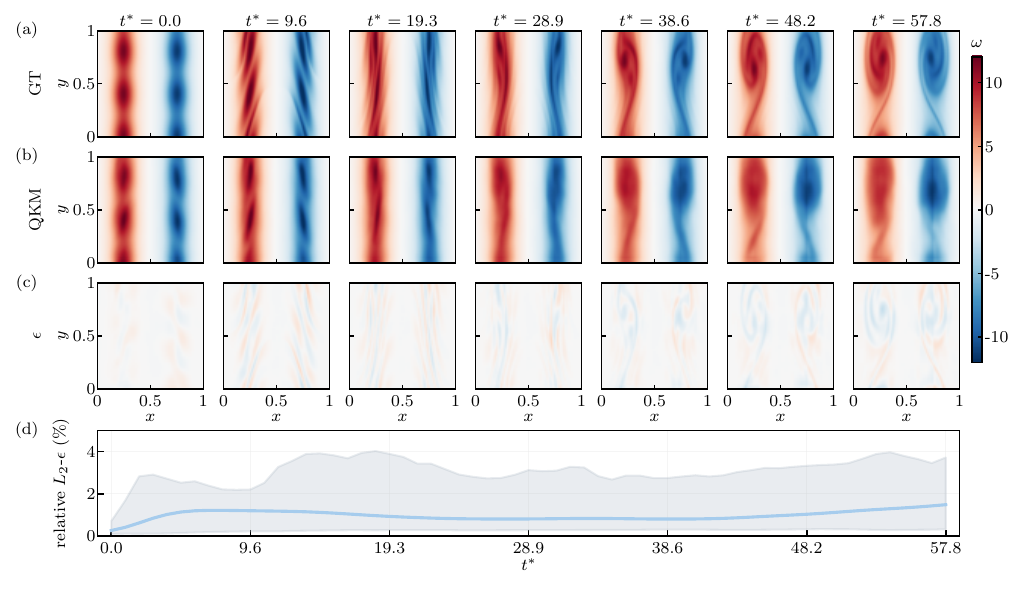}
    \caption{
    Vorticity $\omega$ over $T = 60$ time steps in a representative shear flow case. 
    (a) GT, (b) QKM predictions, (c) Pointwise absolute error $\epsilon(x,y;t^{*}) = \bigl|\omega^{\mathrm{QKM}}(x,y;t^{*}) - \omega^{\mathrm{GT}}(x,y;t^{*})\bigr|$, and (d) Relative $L_2$ error over time for this case (solid line) with error band representing the 10th to 90th percentile of values across all test shear flow cases.
    }
    \label{fig:shear_contour}
\end{figure}

\begin{figure}[!ht]
    \centering
    \includegraphics[width=1.0\linewidth]{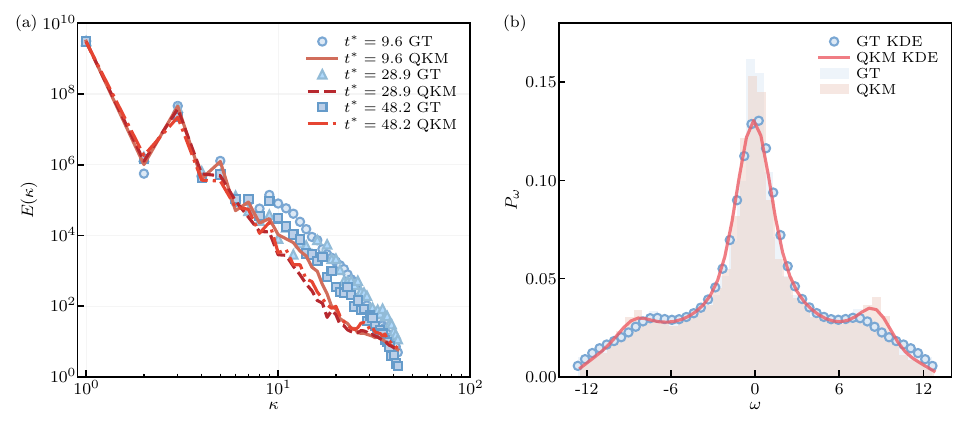}
    \caption{
    Statistics of the shear flow case. 
    (a) Energy spectra \(E(\kappa)\) comparing GT (blue markers) and QKM at \(t^*=9.6\) (solid red), \(t^*=28.9\) (dashed red), and \(t^*=48.2\) (dash–dot red).
    (b) PDF \(P_\omega\) at \(t^{*}=9.6\) comparing histograms (bars) and KDE (lines) for GT (blue) and QKM (red).
    }
    \label{fig:shear_stat}
\end{figure}

Figure~\ref{fig:shear_contour} plots QKM's prediction of vorticity over 60 time steps for a representative shear flow case.
The GT contours in Fig.~\ref{fig:shear_contour}(a) show characteristic rollup of vortical structures from the initial shear layer.
The QKM predictions in Fig.~\ref{fig:shear_contour}(b) reproduce the vortex rollup patterns and shear layer evolution, maintaining structural fidelity across all time steps.
The pointwise absolute error fields in Fig.~\ref{fig:shear_contour}(c) reveal low-magnitude discrepancies.
These errors are primarily localized near vortex cores and shear interfaces where velocity gradients are steepest, while bulk flow regions remain accurate.
Figure~\ref{fig:shear_contour}(d) shows the relative $L_2$ error evolution of this case, which remains below 2\%.
QKM exhibits consistent predictive accuracy across all shear flow test cases. 
The relative $L_2$ error distribution shows that 80\% of predictions fall within 0\% to 4\% between the 10th and 90th percentiles, as illustrated in Fig.~\ref{fig:shear_stat}(d).

In Fig.~\ref{fig:shear_stat}(a), the energy spectra $E(\kappa)$ comparison across three temporal snapshots at $t^*=$ 9.6, 28.9, and 48.2 reveals that QKM maintains good agreement with GT at low to moderate wavenumbers, capturing the primary energy-containing scales of the flow. 
However, notable discrepancies emerge at high wavenumbers, where QKM exhibits energy deficits compared to GT. 
This behavior aligns consistently with the spatial error patterns observed in Fig.~\ref{fig:shear_contour}.
This reflects the challenge in accurately predicting the nonlinear energy cascade across all scales, where energy transfer from large to small scales involves complex mode interactions.
The vorticity PDF in Fig.~\ref{fig:shear_stat}(b) demonstrates QKM's robust statistical fidelity. 
Both the histogram and KDE show correspondence between predicted and GT vorticity statistics, indicating that QKM preserves the fundamental statistical characteristics of the turbulent shear flow despite the localized high-wavenumber discrepancies. 

\subsection{2D turbulence}

The forced 2D turbulence is governed by the incompressible NS equations in Eq.~\eqref{eq:NS}, where we consider a specific forcing term $\bm{f} = \sin{(4y)}\bm{e}_x - 0.1\bm{u}$. 
The forcing combines a sinusoidal profile $\sin(4y)\bm{e}_x$ in the $x$-direction and a linear damping $-0.1\bm{u}$ that prevents large-scale energy accumulation~\cite{boffetta2012two}. 
This maintains statistically stationary turbulence with controlled energy injection, where the flow complexity depends solely on $\Rey$.
The dimensionless time is defined as $t^* = t \langle |\omega_0|_{\mr{max}} \rangle$.

The data was obtained from Ref.~\cite{kochkov2021machine} with $\Rey=\num{4000}$. 
The computational domain spans $x,y \in [0,4\pi]$ using $128^2$ uniform grid points. 
The dataset contains $\num{1680}$ trajectories, each consists of $\num{61}$ temporal snapshots separated by \( \Delta t = 0.1 \), sampled consecutively from the 1286-step solutions.

\begin{figure}[!ht]
    \centering
    \includegraphics[width=1.0\linewidth]{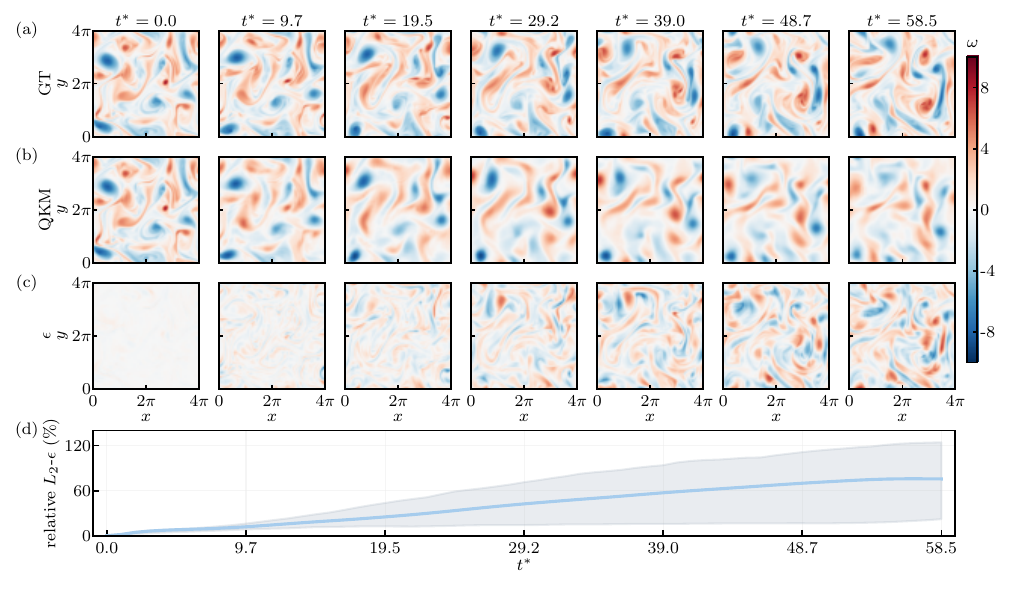}
    \caption{
    Vorticity $\omega$ over $T = 60$ time steps in a representative 2D turbulence case. 
    (a) GT, (b) QKM predictions, (c) Pointwise absolute error $\epsilon(x,y;t^{*}) = \bigl|\omega^{\mathrm{QKM}}(x,y;t^{*}) - \omega^{\mathrm{GT}}(x,y;t^{*})\bigr|$, and (d) Relative $L_2$ error over time for this case (solid line) with error band representing the 10th to 90th percentile of values across all test shear flow cases.
    }
    \label{fig:kol_contour}
\end{figure}

Figure~\ref{fig:kol_contour} compares QKM predictions with GT on the vorticity evolution.
GT contours in Fig.~\ref{fig:kol_contour}(a) show the complex vortical structure evolution.
QKM predictions in Fig.~\ref{fig:kol_contour}(b) resolve the large-scale structures but lose fine-scale details through evolution.
The absolute and relative $L_2$ errors in Figs.~\ref{fig:kol_contour}(c) and (d) show gradual prediction divergence from GT. 
By $t^* = 58.5$, the relative $L_2$ error for this case reaches 75.9\%, while the 10th to 90th percentiles span from 22.5\% to 124\% for all 2D turbulence cases.
QKM generalization over 60 steps is presented in \ref{app:extrapolation}.
The small-scale energy deficit reflects limitations using finite-dimensional Koopman approximation on strongly nonlinear systems. 
This is further intensified by machine learning's bias toward low-frequency features \cite{Xu2025Overview}, which smooths high-frequency structures. 

\begin{figure}[!ht]
    \centering
    \includegraphics[width=1.0\linewidth]{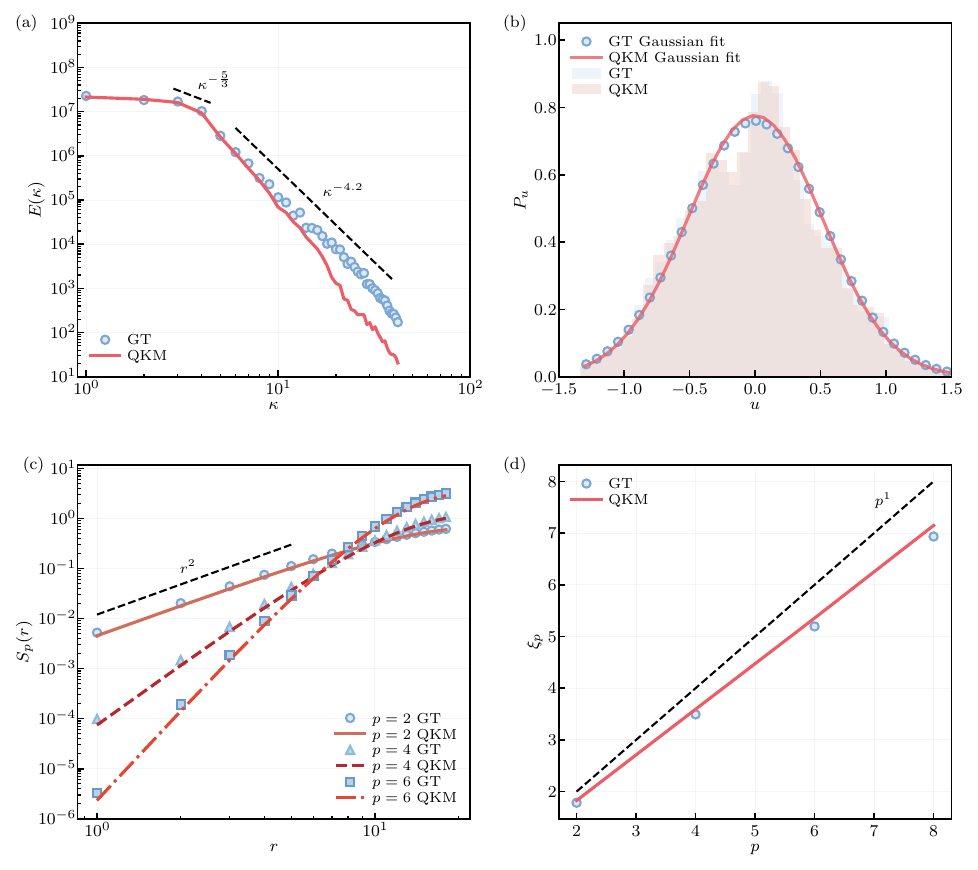}
    \caption{
    Statistics of the 2D turbulence case at $t^* = 9.7$. 
    (a) Energy spectrum showing low-wavenumber $-5/3$ scaling and high-wavenumber $\kappa^{-4.3}$ scaling; 
    (b) Velocity magnitude PDF \(P_{u}\) comparing histograms (bars) and Gaussian fit (lines) for GT (blue) and QKM (red);
    (c) Velocity structure functions $S_p(r)$ for different orders $p$; 
    (d) Scaling exponents $\xi_p$ of velocity structure function versus order $p$.
    }
    \label{fig:kol_stat}
\end{figure}

Figure~\ref{fig:kol_stat} presents key statistics for the 2D turbulence case at $t^{*}=9.7$. 
The energy spectrum in Fig.~\ref{fig:kol_stat}(a) shows excellent agreement between QKM and GT for low and moderate wavenumbers with $\kappa < 10$, both exhibit the characteristic $\kappa^{-5/3}$ scaling.
However, at high wavenumbers with $\kappa > 10$, QKM predictions show lower energy compared to GT, similar to filtering of small-scale turbulence structures. 
Figure~\ref{fig:kol_stat}(b) demonstrates good agreement between QKM and GT on velocity distributions, both closely approximating Gaussian profiles.  
The higher-order velocity structure functions in Fig.~\ref{fig:kol_stat}(c) and their scaling exponents in Fig.~\ref{fig:kol_stat}(d) show that QKM accurately captures turbulent scaling laws at resolved scales. 
These statistics confirm that while QKM preserves essential large-scale turbulent characteristics.

While finite-dimensional linear spaces may adequately represent weakly nonlinear dynamics, 2D turbulence presents a multi-scale coupled strongly nonlinear problem.
The complex energy cascades and nonlinear interactions across multiple scales demand a more comprehensive representation than current approximation provides, requiring higher-dimensional observable spaces. 
These limitations are expected to diminish as model dimensionality increases and autoencoder architectures improve to better preserve fine-scale features. 
\ref{app:size} quantifies these model capacity effects, demonstrating that increased dimensionality reduces validation loss nearly tenfold in reaction-diffusion and shear flow cases, while the scaling behavior for turbulence requires further exploration.
Despite the current constraints, the linear evolution operator architecture presents compelling advantages for quantum computing implementation, with complexity scaling logarithmically with the observable space dimension.  
Future enhancements will focus on scaling network dimensions and implementing high-frequency constraints to strengthen representational capacity, advancing toward quantum evolution modeling of strongly nonlinear dynamics.

\section{Conclusions}\label{sec:conclusion}

We present the data-driven QKM, a framework that bridges nonlinear dynamics and quantum computation.
By leveraging the Koopman operator theory, QKM transforms the fundamental challenge of simulating nonlinear dynamics via global linearization. 
The method employs a hierarchical structure that maps system states into multiple finite-dimensional Hilbert spaces through learned autoencoder networks.
Decomposing system representations into modulus and phase components, the method enables learning unitary operators that govern temporal evolution on the phase space, ensuring compatibility with quantum computation. 
This global linearization enables direct multi-step predictions without iteration, avoiding repetitive measurement and preparation of the quantum state. 

The globally linearized system is governed by a block-diagonal unitary operator, where each subsystem is approximated as an ergodic rotation on a torus. 
This approximation yields unitary operators as tensor products of single-qubit rotations, enabling implementation through $n_j$ parallel $R_z$ gates for each subsystem of $n_j$ qubits. 
Consequently, computational complexity of QKM scales as $\mc{O}(\log N_j)$ for the observable space dimension $N_j = 2^{n_j}$, compared to $\mc{O}(N_j)$ complexity for classical Koopman implementations.  
Furthermore, the quantum circuit structure is hardware-friendly by avoiding entangling gates. 
Training QKM employs a memory-efficient loss function combining reconstruction and prediction components, jointly optimizing the autoencoder mappings and unitary evolution operators through supervised learning on time-series data.

Validation experiments across three nonlinear systems demonstrate varying degrees of success. 
For reaction-diffusion dynamics, QKM accurately reproduces pattern evolution and statistical properties, maintaining relative $L_2$ errors below 6\% across most test cases. 
Shear flow simulations show relative errors within 4\% for the test cases, with good preservation of large-scale vortical structures and energy distributions. 
However, performance degrades for 2D turbulence, where complex multi-scale nonlinear interactions challenge the finite-dimensional linear approximation. 
While large-scale statistics and energy scaling laws are captured accurately, high-wavenumber energy deficits indicate limitations in representing strongly nonlinear dynamics.
The parameter analysis reveals that model capacity significantly influences performance, with increased subsystem dimensions reducing validation losses substantially for weakly nonlinear systems. 
Extrapolation tests confirm the method's ability to generalize beyond training time steps, though accuracy depends on the degree of nonlinearity in the underlying dynamics.

Several limitations constrain the current approach. 
The finite-dimensional linear approximation proves insufficient for strongly nonlinear systems with complex energy cascade processes, while neural network bias toward low-frequency features intensifies the loss of high-frequency information.  
Additionally, the decomposition strategy and hierarchical encoding may not optimally represent all classes of dynamical systems.
The current framework lacks theoretical guarantees on the relationship between prediction accuracy and embedding dimension. 
Although our empirical results demonstrate effectiveness with polynomial dimension increases for the tested systems, theoretical analysis is needed to establish fundamental scaling limits.
Future developments should enhance representational capacity through higher-dimensional observable spaces and architectural improvements to preserve fine-scale nonlinear features. 
Incorporating explicit high-frequency component modeling \cite{bai2022improving}, physics-informed loss~\cite{Raissi2018Hidden,Wang2017Physicsinformed,zhao2025lesnets}, and alternative encoding strategies may yield better approximation and enhance small-scale resolution. 

The QKM establishes a pathway for quantum simulation of nonlinear dynamics by exploiting the synergy between machine learning for global linearization and quantum computing for unitary evolution. 
Although QKM's simple quantum circuit architecture is hardware-friendly for NISQ devices, critical challenges remain for efficient state preparation and measurement~\cite{Aaronson2015Read,Aaronson2025Future}, when implementing on real quantum computers. 
Decoherence effects and quantum noise can corrupt the learned dynamics, requiring robust error correction/mitigation strategies tailored to the specific circuit structure. 

\section*{Declaration of competing interest}
The authors declare no conflict of interest. 

\section*{Acknowledgments}

The authors thank Z. Meng for helpful suggestions. 
This work has been supported in part by the National Natural Science Foundation of China (Nos.~12525201, 52306126, 12432010, 12588201), the National Key R\&D Program of China (Grant No.~2023YFB4502600), and the Xplore Prize.

\section*{Data availability}

The open source code release is provided in \url{https://github.com/YYgroup/QKoopman}.
The dataset is available at \url{https://huggingface.co/datasets/YYgroup/QKoopman}.

\appendix

\setcounter{figure}{0}
\setcounter{table}{0}

\section{Autoencoder of Koopman observables}\label{app:ae}

The autoencoder architecture serves as the critical interface between the original state space and the hierarchical quantum representation, learning to encode input fields into modulus and phase components across subsystems and reconstruct evolved states from the observables. 
The hierarchical encoding transforms the input state $\vx$ into a sequence of $h$ subsystems through progressively refined feature extraction. 
To ensure functional consistency, modulus $f_r\lrr{\vx}$ and phase $f_\phi\lrr{\vx}$ encoders share identical mathematical formulations and backbone architecture. 
Take the modulus encoder $f_r\lrr{\vx}$ as example, the encoding proceeds as
\begin{align}
    \vr_{1} &=  \mathcal{R}\lrs{\vx\lrr{0}}, \nonumber \\
    \vr_{2} &=  \mathcal{C}\lrc{\mathcal{D}(\vr_1) + \mathcal{R} \circ \mathcal{C}\lrs{\vx\lrr{0}}}, \label{eq:encoder} \\
    \vr_{j} &= \mathcal{C}\lrc{
        \mathcal{D}\lrr{\vr_{j-1}} + 
        \underbrace{\mathcal{T} \circ \mathcal{T} \circ \cdots \circ \mathcal{T}}_{(j-2)\text{ times}} 
        \circ \mathcal{R} \circ \mathcal{C}\lrs{\vx\lrr{0}}
    }
    \quad \text{for} \quad j = 3,4,\ldots,h, \nonumber
\end{align}
where $\mathcal{R}(\cdot)$, $\mathcal{C}(\cdot)$, and $\mathcal{T}(\cdot)$ denote the residual, convolution, and transformer blocks, respectively, 
$\mathcal{D}(\cdot)$ represents a $3 \times 3$ convolution operator with stride $2$ for downsampling. 

The decoder reconstructs the evolved state from the quantum-processed observables, mapping the modulus $\vr$ and evolved phase $\bm{\phi}$ back to the state space $\mc{X}$ through the learned inverse transformation
\begin{align}
    \vy_{h} &= \vr_{h}\cos{\bm{\phi}_h\lrr{t}}, \nonumber \\
    \vy_{j} &= \mathcal{R} \circ \mathcal{C} \lrc{w\lrs{\vr_{j}\cos{\bm{\phi}_j\lrr{t}}, \mathcal{P}\lrr{\vy_{j+1}}}} \quad \text{for} \quad j = 2,\ldots,h-1,  \label{eq:decoder} \\
    \vx\lrr{t} &= \mathcal{Q} \circ \mathcal{C} \lrc{w\lrs{\vr_{1}\cos{\bm{\phi}_1\lrr{t}}, \mathcal{P}\lrr{\vy_{2}}}}, \nonumber
\end{align}
where $\mc{P}(\cdot)$ denotes a $3\times 3$ transposed convolution with stride 2 for upsampling, 
$\mc{Q}(\cdot)$ is a $1\times 1$ convolution with stride 1 for output refinement,  
and $w(\cdot,\cdot)$ handles tensor reshaping and channel-wise concatenation. 

\begin{figure}[!ht]
    \centering
    \includegraphics[width=\textwidth]{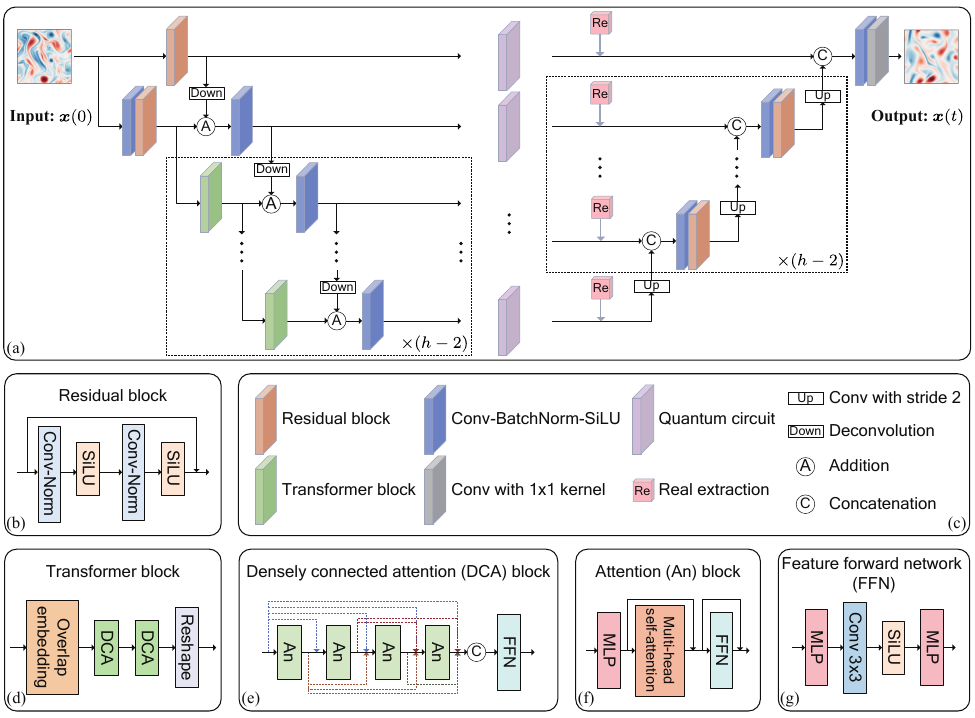}
    \caption{Architecture of the Koopman observable autoencoder. 
    (a) Overall pipeline showing the hierarchical encoding of input fields into subsystems through progressive downsampling operations, quantum evolution processing, and hierarchical decoding with upsampling to reconstruct evolved states. 
    (b) Residual block architecture integrating dual Conv--BatchNorm--SiLU layers.  
    (c) Architectural symbol legend defining computational operations including convolution, deconvolution, batch normalization, activation functions, and tensor operations. 
    (d) Transformer block structure employing overlapping patch embedding followed by DCA units. 
    (e) DCA block cascading multiple attention modules with an FFN.  
    (f) Attention module implementing multi-head self-attention mechanisms through MLP-based transformations. 
    (g) FFN performing channel mixing and feature transformation through cascaded Conv--SiLU--MLP operations. 
    }
    \label{fig:net_detail}
\end{figure}

The autoencoder is realized through a modified U-Net~\cite{ronneberger2015u} architecture enhanced with vision transformer components~\cite{li2025transformer}, as illustrated in Fig.~\ref{fig:net_detail}. 
The system employs three distinct computational blocks for encoding and their corresponding inverse operations for decoding, with skip connections facilitating information flow between corresponding encoder-decoder levels.
It combines convolutional neural networks for spatial processing and transformer mechanisms for long-range dependencies, forming a framework for learning the global linearized observable space of dynamical systems.

The residual blocks $\mc{R}\lrr{\cdot}$ in Fig.~\ref{fig:net_detail}(b) integrate dual convolution (Conv)--batch normalization (BatchNorm)--sigmoid linear unit (SiLU) layers with skip connections to facilitate gradient flow and preserve fine-grained features. 
The convolution blocks $\mc{C}\lrr{\cdot}$ perform spatial downsampling during encoding and feature processing during decoding. 
The transformer blocks $\mc{T}\lrr{\cdot}$ in Fig.~\ref{fig:net_detail}(d) utilize overlap embedding~\cite{xie2021segformer} with densely connected attention (DCA)~\cite{shi2023h} units in Fig.~\ref{fig:net_detail}(e) to capture long-range spatial dependencies in the feature maps.
The DCA blocks in Fig.~\ref{fig:net_detail}(e) cascade multiple attention modules with feed-forward networks (FFN), enabling rich feature interactions across spatial locations. 
Each attention module in Fig.\ref{fig:net_detail}(f) implements multi-head self-attention mechanisms through multi-layer perceptron (MLP)-based transformations for learning spatial relationships in the encoded features. 
The FFNs in Fig.~\ref{fig:net_detail}(g) perform channel mixing and feature transformation through cascaded Conv--SiLU--MLP operations to enhance representational capacity.

The decoder mirrors the encoder's hierarchical structure but operates in reverse, progressively upsampling through transposed convolutions $\mc{P}\lrr{\cdot}$ and integrating features across scales. 
The modulus-phase decomposition $\vr_j \cos \bm{\phi}_j \lrr{t}$ at each level directly incorporates quantum evolution results, while concatenation operations  ensure proper feature fusion across the autoencoder's latent space.

\setcounter{figure}{0}
\setcounter{table}{0}

\section{Long-term prediction}\label{app:extrapolation}

To evaluate the model's generalization beyond its training time steps and robustness in capturing long-term nonlinear dynamics, we test its performance on temporal extrapolation. 
The model, trained on 61-step sequences, forecasts the subsequent 10 time steps on unseen test trajectories. 
This extrapolation test assesses the model's ability to maintain predictive accuracy outside its training domain, a critical requirement for simulating complex dynamical systems. 
The results below demonstrate performance on this challenging task.

\begin{figure}[!ht]
    \centering
    \includegraphics[width=1.0\linewidth]{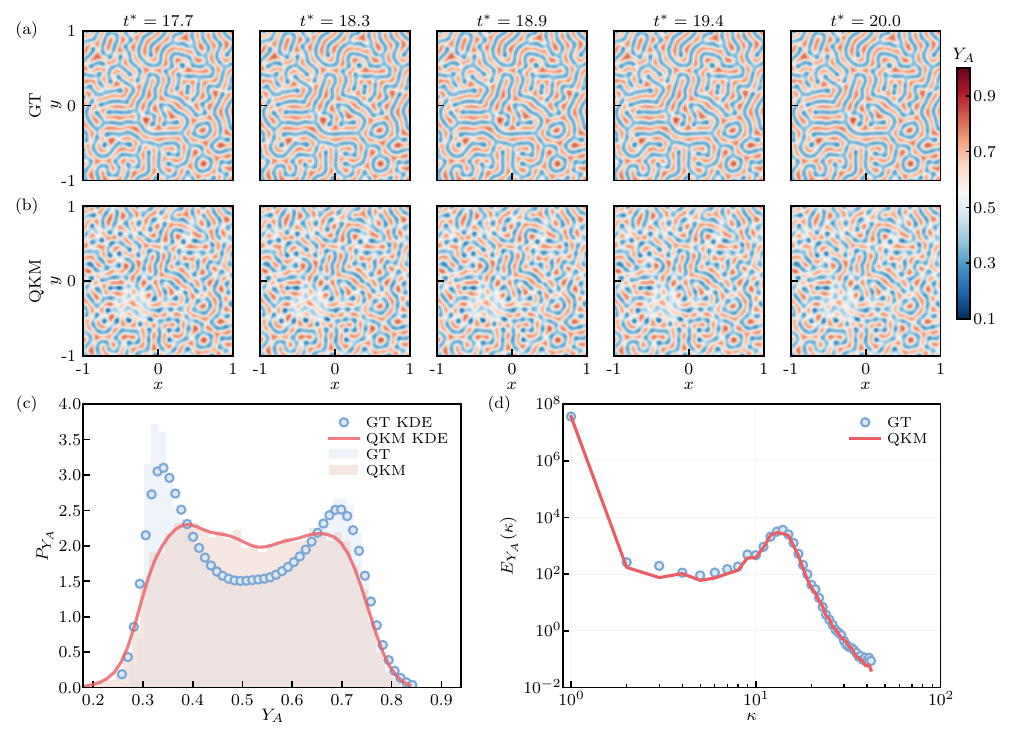}
    \caption{
    Species concentration $Y_A$ for the reaction-diffusion system with $(F, K) = (0.029, 0.057)$ from 61 to 70 steps.
    (a) GT, (b) QKM prediction, and (c) $P_{Y_A}$ and (d) $E_{Y_A}(\kappa)$ at $t^{*}=18.9$. }
    \label{fig:case2ex}
\end{figure}

Figure~\ref{fig:case2ex}(a) and (b) compare contours from GT and QKM at extrapolated time steps for the Gray-Scott reaction-diffusion system shown in Fig.~\ref{fig:gray_contour}. 
QKM reproduces the dominant spatial pattern, including filamentary structures and domain-scale organization, with reasonable visual agreement across all extrapolated steps. 
The relative $L_2$ error remains low throughout extrapolation, increasing gradually from 1.43\% at $t^* = 17.7$ to 1.77\% at $t^* = 20.3$, indicating stable predictive behavior.

Figure~\ref{fig:case2ex}(c) presents $P_{Y_A}$ at $t^*=18.9$. 
While histogram shapes are similar, the KDE reveals discrepancies in capturing the bimodal character of the GT distribution, suggesting QKM reproduces coarse statistical properties but has limitations in resolving finer distributional details. 
Figure~\ref{fig:case2ex}(d) shows the spectrum $E_{Y_A}$ at the same time, where QKM closely follows GT across a broad range of spatial frequencies, indicating the model retains dominant spatial scales and captures multi-scale structure.
These results demonstrate that QKM maintains reasonable extrapolation accuracy for the reaction-diffusion case.

\begin{figure}[!ht]
    \centering
    \includegraphics[width=1.0\linewidth]{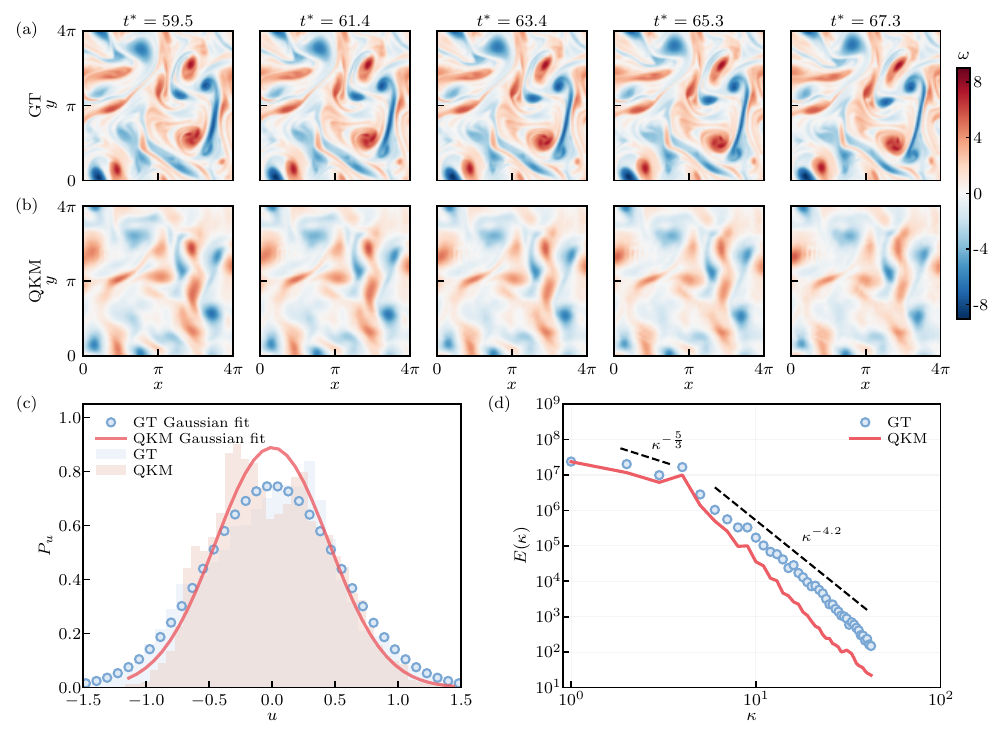}
    \caption{
    Flow fields for the 2D turbulence from 61 to 70 steps.
    Vorticity $\omega$ at $t^* = 59.5$-$68.2$ of (a) GT and (b) QKM predictions, 
    (c) Velocity PDF $P_u$ at $t^* = 63.4$, comparing histograms (bars) and Gaussian fits (lines), 
    (d) Energy spectrum $E(\kappa)$ at $t^* = 63.4$.
    }
    \label{fig:case1ex}
\end{figure}

Figure~\ref{fig:case1ex} illustrates QKM's extrapolation performance on 2D turbulence. 
Figure~\ref{fig:case1ex}(a) and (b) compare vorticity fields from GT and QKM. 
QKM recovers large-scale vortices, though the fields appear slightly smoother with underpredicted fine-scale intensity.
Figure~\ref{fig:case1ex}(c) shows the velocity PDF at $t^* = 63.4$, where both GT and QKM distributions are close to Gaussian.
QKM reproduces the overall PDF shape.
Figure~\ref{fig:case1ex}(d) shows the velocity energy spectrum at the same time step. 
QKM perform well at low wavenumbers, but shows discrepancies at high wavenumbers, similar to the results in Fig.~\ref{fig:kol_contour}.

\setcounter{figure}{0}
\setcounter{table}{0}

\section{Ablation study on model architecture parameters}\label{app:size}

We investigate the effects of finite-dimensional approximations through two parameters: the subsystem count $h$ and the feature channel count $c$. 
Our ablation study varies these parameters systematically while maintaining all other experimental settings identical.

\begin{figure}[!ht]
    \centering
    \includegraphics[width=1.0\linewidth]{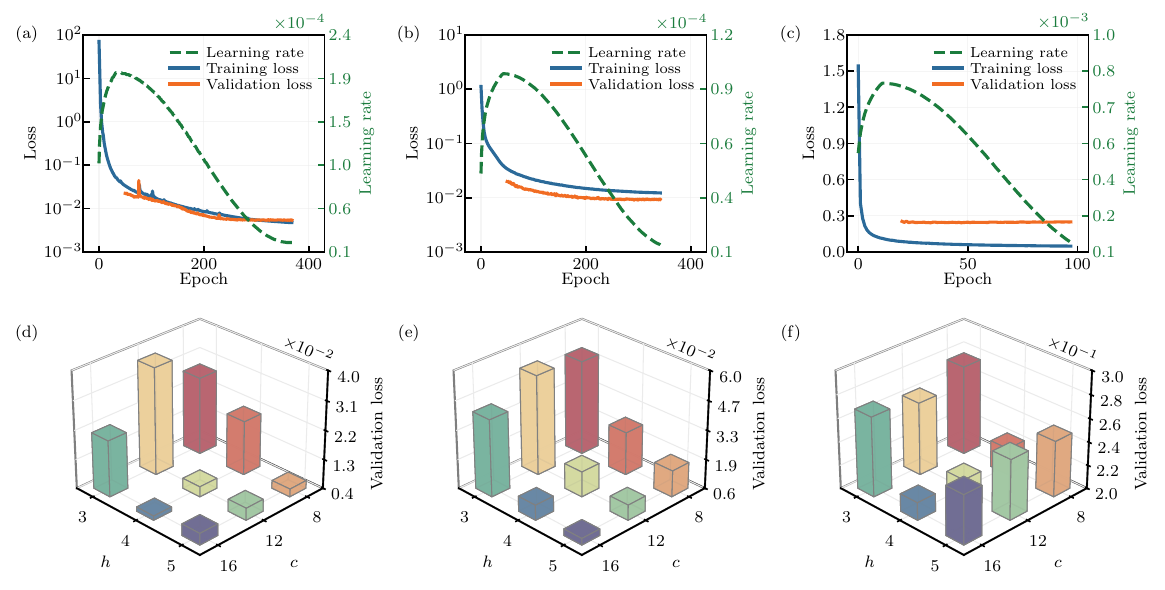}
    \caption{QKM training dynamics and performance: 
    (a-c) Training and validation losses evolution with learning rate schedules for (a) reaction-diffusion, (b) shear flow, and (c) 2D turbulence;
    (d-f) Final validation losses across model configurations ($h = 3,4,5$ and $c = 8,12,16$) for each respective case.
    }
    \label{fig:3drmse}
\end{figure}

Figure~\ref{fig:3drmse}(a--c) show the evolution of training and validation losses along with the learning rate schedule for the reaction-diffusion system, shear flow, and 2D turbulence cases.
Validation losses decrease steadily with convergence behavior.
Figure~\ref{fig:3drmse}(d--f) show final validation losses across model configurations where $h$ varies among $\{3, 4, 5\}$ and $c$ among $\{8, 12, 16\}$. 
Although a larger model capacity (i.e., higher $h$ and $c$) generally improves performance, results do not strictly follow this trend.
This is likely due to random initialization effects or limited training data.

Table~\ref{tab:memory} presents QKM computational characteristics across configurations. 
As subsystem count $h$ increases from 3 to 5 and feature channel count $c$ from 8 to 16, model parameters scale from 0.46 million to 11.07 million. 
At batch size $B=1$, inference memory ranges from 68.25 MB for the smallest configuration with $h=3$ and $c=8$ to 145.25 MB for the largest with $h=5$ and $c=16$. 
When scaled to maximum physical batch size $B=240$, memory consumption varies minimally from 16.00 GB to 16.95 GB across all configurations.

\begin{table}[htbp]
\centering
\caption{Model configuration and resource requirements for QKM implementations. Parameters include subsystem count ($h$), feature channel dimension ($c$), total parameters (million), and inference memory for batch sizes $B=1$ (MB) and $B=240$ (GB).}
\label{tab:memory}
\begin{tabular}{ccccc}
\toprule
Subsystems & Channels & Parameters & Inference memory & Inference memory\\
($h$) & ($c$) & (million) & ($B=1$, MB) & ($B=240$, GB) \\ 
\midrule
\multirow{3}{*}{3} 
    & 8 & 0.46 & 68.25 & 16.00\\
    & 12 & 1.02 & 133.63 & 16.45\\
    & 16 & 1.79 & 135.44 & 16.91\\
\cmidrule(lr){1-5}

\multirow{3}{*}{4}
    & 8 & 1.17 & 131.5 & 16.01 \\
    & 12 & 2.59 & 134.47 & 16.47\\
    & 16 & 4.57 & 136.56 & 16.93\\
\cmidrule(lr){1-5}

\multirow{3}{*}{5}
    & 8 & 2.81 & 131.78 & 16.02 \\
    & 12 & 6.26 & 134.89 & 16.48 \\
    & 16 & 11.07 & 145.25 & 16.95 \\
\bottomrule
\end{tabular}
\end{table}

\bibliography{mycite}
\end{document}